\documentclass[twocolumn,preprintnumbers,amsmath,amssymb,floatfix, superscriptaddress]{revtex4-1}

\usepackage{graphicx}
\usepackage{dcolumn}
\usepackage{bm,epsfig}
\usepackage{epstopdf}
\usepackage[percent]{overpic}

\begin{document}

\title{Electronic and magnetic properties of stoichiometric CeAuBi$_{2}$}

\author{M. M. Piva}
\email{Mario.Piva@cpfs.mpg.de}
\affiliation{Instituto de F\'{\i}sica ``Gleb Wataghin'', UNICAMP, 13083-859, Campinas, SP, Brazil}
\affiliation{Los Alamos National Laboratory, Los Alamos, New Mexico 87545, USA}
\affiliation{Max Planck Institute for Chemical Physics of Solids, N\"{o}thnitzer Str.\ 40, D-01187 Dresden, Germany}

\author{R. Tartaglia}
\affiliation{Instituto de F\'{\i}sica ``Gleb Wataghin'', UNICAMP, 13083-859, Campinas, SP, Brazil}

\author{G. S. Freitas}
\affiliation{Instituto de F\'{\i}sica ``Gleb Wataghin'', UNICAMP, 13083-859, Campinas, SP, Brazil}

\author{J. C. Souza}
\affiliation{Instituto de F\'{\i}sica ``Gleb Wataghin'', UNICAMP, 13083-859, Campinas, SP, Brazil}

\author{D. S. Christovam}
\affiliation{Instituto de F\'{\i}sica ``Gleb Wataghin'', UNICAMP, 13083-859, Campinas, SP, Brazil}

\author{S. M. Thomas}
\affiliation{Los Alamos National Laboratory, Los Alamos, New Mexico 87545, USA}

\author{J. B. Le\~ao}
\affiliation{NIST Center for Neutron Research, National Institute of Standards and Technology, Gaithersburg, Maryland 20899, USA}

\author{W. Ratcliff}
\affiliation{NIST Center for Neutron Research, National Institute of Standards and Technology, Gaithersburg, Maryland 20899, USA}

\author{J. W. Lynn}
\affiliation{NIST Center for Neutron Research, National Institute of Standards and Technology, Gaithersburg, Maryland 20899, USA}

\author{C. Lane}
\affiliation{Los Alamos National Laboratory, Los Alamos, New Mexico 87545, USA}

\author{J.-X. Zhu}
\affiliation{Los Alamos National Laboratory, Los Alamos, New Mexico 87545, USA}

\author{J. D. Thompson}
\affiliation{Los Alamos National Laboratory, Los Alamos, New Mexico 87545, USA}

\author{P. F. S. Rosa}
\affiliation{Los Alamos National Laboratory, Los Alamos, New Mexico 87545, USA}

\author{C. Adriano}
\affiliation{Instituto de F\'{\i}sica ``Gleb Wataghin'', UNICAMP, 13083-859, Campinas, SP, Brazil}

\author{E. Granado}
\affiliation{Instituto de F\'{\i}sica ``Gleb Wataghin'', UNICAMP, 13083-859, Campinas, SP, Brazil}

\author{P. G. Pagliuso}
\affiliation{Instituto de F\'{\i}sica ``Gleb Wataghin'', UNICAMP, 13083-859, Campinas, SP, Brazil}

\date{\today}

\begin{abstract}

We report the electronic and magnetic properties of stoichiometric CeAuBi$_{2}$ single crystals. At ambient pressure, CeAuBi$_{2}$ orders antiferromagnetically below a N\'{e}el temperature ($T_{N}$) of 19~K. Neutron diffraction experiments revealed an antiferromagnetic propagation vector $\hat{\tau} = [0, 0, 1/2]$, which doubles the paramagnetic unit cell along the $c$-axis. At low temperatures several metamagnetic transitions are induced by the application of fields parallel to the $c$-axis, suggesting that the magnetic structure of CeAuBi$_{2}$ changes as a function of field.  At low temperatures, a linear positive magnetoresistance may indicate the presence of band crossings near the Fermi level. Finally, the application of external pressure favors the antiferromagnetic state, indicating that the 4$f$ electrons become more localized.

\end{abstract}

\maketitle

\section{INTRODUCTION}

Materials with topological non-trivial phases are being extensively studied due to their potential in enabling new technologies \cite{Tokura,Darkmatter,Spintronics}. In particular, topological semimetals are predicted to host band crossings whose low-energy excitations mimic relativist (Dirac or Weyl) fermions. Experimentally, these materials exhibit extremely large magnetoresistance, ultrahigh mobilities, and intrinsic anomalous Hall effect \cite{RevTopo, AHE}. In this regard, nonsymmorphic crystalline structures have been recently predicted to naturally give rise to symmetry-protected band crossings \cite{Nonsym}. In particular compounds that crystallize in the $P4/nmm$ structure with square nets are predicted to host band crossings at $M$, $X$, $A$ and $R$ points of the Brillouin zone, even in the presence of spin-orbit coupling \cite{Squarenet}. This prediction was confirmed by ARPES measurements on ZrSiS and HfSiS, which crystallize in the $P4/nmm$ structure and present Dirac line nodes in their electronic band structures \cite{ARPESSiS1, ARPESSiS2}.

Many materials crystallizing in the $P4/nmm$ structure contain rare-earth elements, which may enable the yet underexplored interplay between magnetism and topology. For instance, in CeSbTe the application of an external magnetic field leads to a rich phase diagram in which the magnetic ordering breaks additional symmetries leading to new topological phases \cite{CeSbTe}.

\begin{figure}[!b]
	\includegraphics[width=0.5\textwidth]{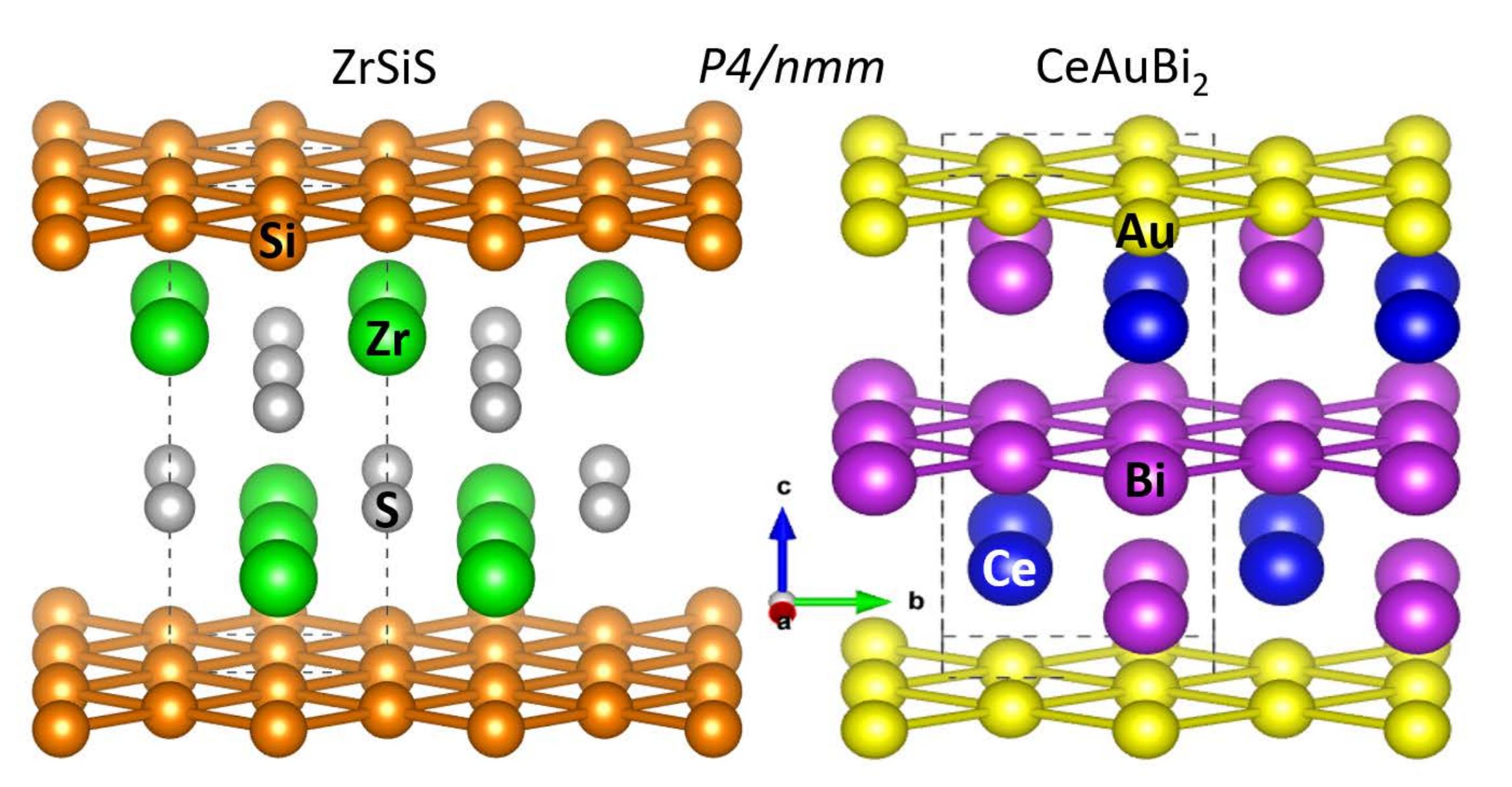}
	\caption{Crystalline structure of ZrSiS and CeAuBi$_{2}$.}
	\label{xtal}
\end{figure}

Here we revisit the compound CeAuBi$_{2}$. It also crystallizes in the $P4/nmm$ structure, but with two square nets, one of Au and one of Bi, instead of just one of Si/Sb as in ZrSiS and CeSbTe, as shown in Fig.~\ref{xtal}. We find that CeAuBi$_{2}$ presents an antiferromagnetic propagation vector (0, 0, 1/2) similar to CeSbTe \cite{CeSbTe}. Both structures present ferromagnetic Ce$^{3+}$ planes, which are antiferromagnetically coupled. However, the stacking of these planes is $++--$ in CeAuBi$_{2}$, whereas in CeSbTe a $+--+$ structure is observed \cite{CeSbTe}.

The synthesis of stoichiometric CeAuBi$_{2}$ single crystals, however, has been missing due to the presence of Au vacancies \cite{CrisAu,CBRAu,Cava}. Here we report the synthesis and characterization of stoichiometric CeAuBi$_{2}$ single crystals for the first time. By preventing Au vacancies we find that the antiferromagnetic transition temperature ($T_{N}$) increases to 19~K, the highest value reported for this compound \cite{CrisAu,CBRAu, Cava}. Moreover, we find that the application of external pressure enhances the antiferromagnetic order, driving $T_{N}$ to 21~K at 23~kbar ($1$~GPa = $10$~kbar).  The high quality of our single crystals also enabled the discovery of several metamagnetic transitions at 2~K and ambient pressure as a function of applied magnetic field parallel to the $c$-axis. The critical fields of these transitions follow $T_{N}$, increasing as a function of applied pressure. At 2~K a linear response of the magnetoresistance (MR) is observed for all studied pressures, which may arise from band crossings near the Fermi level. Also, Hall resistivity ($\rho_{xy}$) measurements at 2~K suggest the presence of multi-band effects that are enhanced by the application of external pressure. Finally, neutron magnetic diffraction measurements at zero field show a magnetic structure that doubles the unit cell along the $c$-axis identical to CeCuBi$_{2}$ \cite{CrisCu}.   

\section{EXPERIMENTAL DETAILS}

Single crystals of CeAuBi$_{2}$ were grown by the Bi-flux technique with starting composition Ce:Au:Bi = 1:1.5:20. The elements were put in an alumina crucible and inside a quartz tube, which was sealed under vacuum. The tube was heated to 850$^{\circ}$C in 8 hours and was kept 12 hours at 850$^{\circ}$C. After this, it was cooled to 550$^{\circ}$C in 100 hours, followed by annealing for one day. The excess of Bi was removed by spinning the tube upside down in a centrifuge. A commercial X-ray diffractometer was used to check the crystallographic structure by single-crystal diffraction at room temperature. Moreover, the composition of the compound was checked by performing energy dispersive X-ray spectroscopy (EDX) on a polished single crystal, yielding CeAu$_{0.95(3)}$Bi$_{2.22(3)}$. The excess of Bi is common in self-flux grown single crystals, as some residual Bi flux on the surface prevents a more accurate EDX measurement. Therefore, our single crystals are very close to being stoichiometric. Magnetization measurements were performed with a commercial platform equipped with a VSM option. Specific heat measurements at ambient pressure were done using the thermal relaxation technique. Electrical resistivity experiments were done in a four-probe configuration along with a low-frequency AC bridge. Pressures up to 23~kbar were generated using a a self-contained double-layer piston-cylinder-type Cu-Be pressure cell with an inner-cylinder of hardened NiCrAl. The pressure transmitting medium used was Daphne oil, and lead served as a manometer. The specific heat experiments under pressure were performed using an AC calorimetry technique \cite{sullivan, AC}.

Neutron magnetic diffraction experiments with incident energy of 14.7 meV were performed on the BT-7 triple axis spectrometer at the NIST Center for Neutron Research (NCNR) \cite{BT7}. A single crystal with dimensions close to 6~mm x 7~mm x 1~mm  was cooled to the base temperature of 2.8~K using a closed cycle refrigerator. The horizontal collimators were open-50-50-120, with pyrolytic graphite monochromator and analyzer. Note that uncertainties where indicated throughout represent one standard deviation.

Band structure calculations were performed within the framework of density functional theory (DFT) with the full-potential linearized augmented plane wave (FP-LAPW) method as implemented in the WIEN2k code \cite{Blaha}. A spin-polarized generalized gradient approximation (GGA) \cite{Perdew} was used. The muffin-tin radii were 3.0~$a_0$ (Ce), 2.5~$a_0$ (Au), and 2.5~$a_0$ (Bi), where $a_0$ is the Bohr radius. The localized Ce-4$f$ electrons were described by the GGA+$U$ method \cite{Anisimov} with $U = 7.0$~eV and $J = 0.69$~eV.

\section{RESULTS}

\begin{figure}[!t]
	\includegraphics[width=0.5\textwidth]{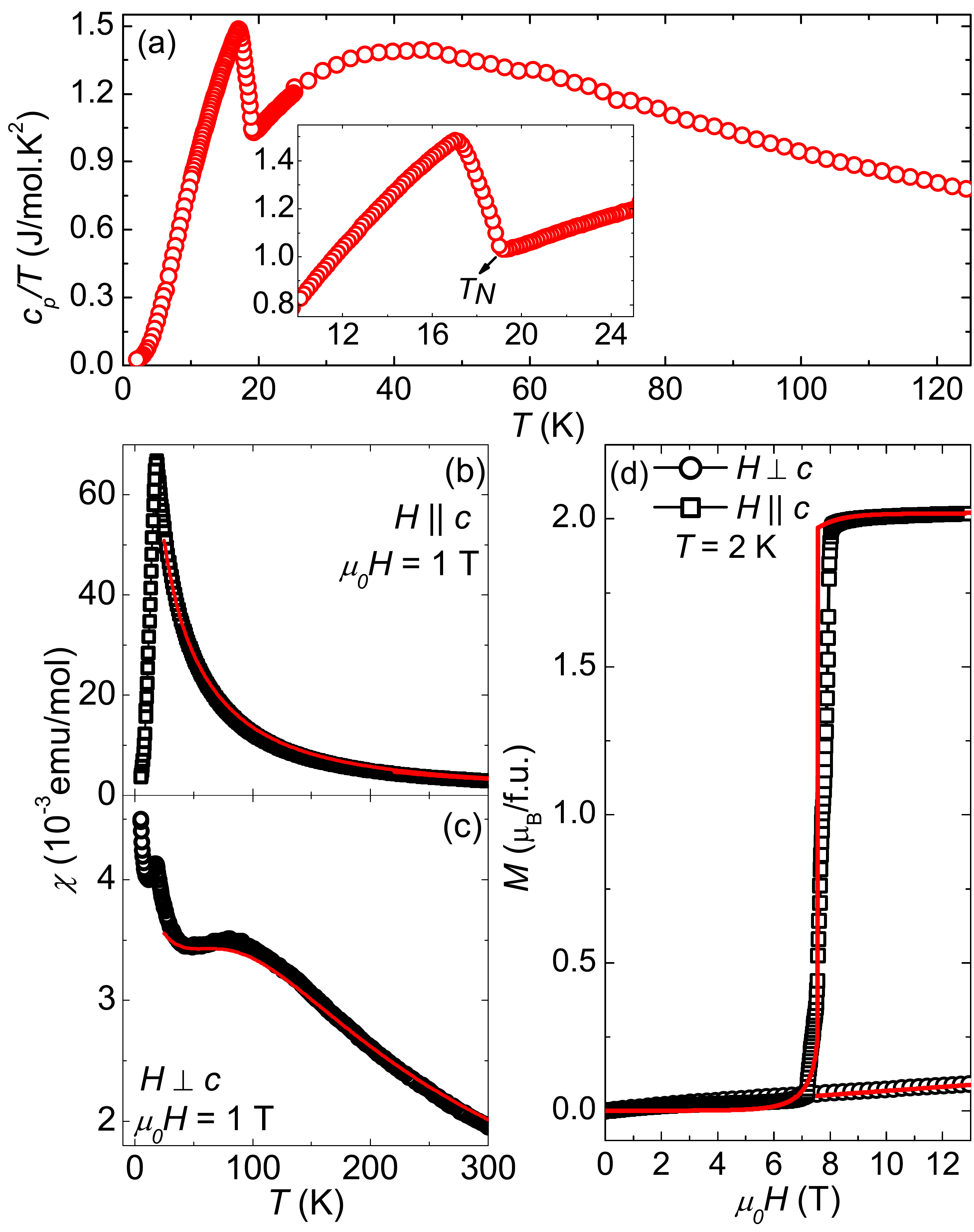}
	\caption{(a) Specific heat divided by temperature as a function of temperature. Magnetic susceptibility as a function of temperature for fields parallel (b) and perpendicular (c) to the $c$-axis. (d) Magnetization as a function of magnetic field at 2~K for fields parallel and perpendicular to the $c$-axis. The solid red lines are fits using a CEF mean-field model. Note that $1$~emu $=$ $10^{-3}$~Am$^{2}$}
	\label{chi}
\end{figure}

At room temperature, CeAuBi$_{2}$ crystallizes in the $P4/nmm$ tetragonal crystal structure, as shown in Fig.~\ref{xtal}, with lattice parameters $a = 4.628(6)$~\AA \- and $c = 9.897(13)$~\AA, in agreement with previous reports \cite{CrisAu, CBRAu, Cava}. 

The specific heat divided by the temperature ($c_{p}/T$)  as a function of temperature at ambient pressure is shown in Fig.~\ref{chi} (a). An upturn at 19~K characterizes the onset of the antiferromagnetic transition, which is followed by a broad peak most likely caused by the presence of residual disorder in the system. The absence of the non-magnetic analog LaAuBi$_{2}$ along with the presence of antiferromagnetic order prevented a reliable estimation of the Sommerfeld coefficient for CeAuBi$_{2}$. Figures~\ref{chi} (b) and (c)  display the response of the magnetic susceptibility ($\chi$) as a function of temperature for CeAuBi$_{2}$ for fields parallel and perpendicular to the $c$-axis, respectively. We extracted Ce$^{3+}$ effective moments of $\mu_{eff} = 2.55(1)$~$\mu_{B}$, by performing Curie-Weiss fits in the high-temperature range of the magnetic susceptibility with fields parallel to the $c$-axis. The obtained $\mu_{eff}$ is identical to the calculated 2.54~$\mu_{B}$ for a free Ce$^{3+}$ ion. At low temperatures, CeAuBi$_{2}$ displays an antiferromagnetic ordering at $T_{N} \approx 19$~K, the highest value reported for this compound. A substantial magnetic anisotropy at $T_{N}$ ($\chi_{//c}/\chi_{\perp c} \approx 16$) is observed, in agreement with previous reports  \cite{CrisAu, CBRAu, Cava}. The broad hump in $\chi$ for $H \perp c$ can be attributed to the first excited crystal-field state. Figure~\ref{chi} (d) presents the magnetization as a function of applied magnetic field at 2~K for both directions. For $H \ || \  c$, a spin-flop transition takes place at around 7.5~T. For $H \perp c$ the magnetization increases monotonically with field. The solid red lines in the main panels of Figs.~\ref{chi} are fits using a crystalline electric Field (CEF) mean field model considering anisotropic nearest-neighbor interactions and the tetragonal CEF Hamiltonian: $\mathcal{H} = g_{J}\mu_{B}\boldsymbol{H}\cdot \boldsymbol{J} + z_{i}J_{i}^{ex} \cdot  \langle J^{ex}\rangle + B^{0}_{2}O^{0}_{2} + B^{0}_{4}O^{0}_{4} + B^{4}_{4}O^{4}_{4}$, where $g_{J}$ is the Land\'e $g$-factor, $\mu_{B}$ is the Bohr magneton, $\boldsymbol{H}$ is the applied magnetic field and $\boldsymbol{J}$ is the total angular momentum. $z_{i}J_{i}^{ex}$ represents the $J_{i}$ mean field interactions ($i = $AFM, FM) between the $z_{i}$ nearest neighbors that mimic the RKKY interaction. $B^{m}_{n}$ are the CEF parameters and the $O^{m}_{n}$ are the Steven's operators \cite{pagliuso2006}. By simultaneously performing fits to $\chi(T)$ and $M(H)$ data above $T_{N}$, we extract the CEF scheme and two RKKY parameters for this compound. For the RKKY parameters we obtain $z_{AFM}J_{AFM}^{ex} = 1.65$~K and $z_{FM}J_{FM}^{ex} = -0.35$~K. The presence of ferromagnetic and antiferromagnetic exchange interactions is in agreement with the transitions observed in CeAuBi$_{2}$ at high fields. For the CEF parameters we obtained the following values: $B^{0}_{2} \approx -17.2$~K, $B^{0}_{4} \approx 0.05$~K and $B^{4}_{4} \approx 0.60$~K. These parameters imply a ground state composed of a $\Gamma_{7}^{1} = 0.99|\pm5/2\rangle - 0.08|\mp3/2\rangle$ doublet, a first excited state $\Gamma_{7}^{2} = 0.08|\mp5/2\rangle + 0.99|\mp3/2\rangle$ doublet at 200~K and a second excited state of a  $\Gamma_{6} = |\pm1/2\rangle$ doublet at 315~K. We note that the CEF parameters acquired from the fits of macroscopic data may not be fully accurate or unique and additional microscopic measurements to confirm this CEF scheme, such as X-ray absorption and/or inelastic neutron scattering, would be desirable. Therefore, this CEF scheme must be treated with caution. Nevertheless removing the Au vacancies of CeAu$_{0.92}$Bi$_{1.6}$ leads to a higher $B^{0}_{2}$ parameter and a higher $T_{N}$. This result reinforces the general trend observed in the 112 family of materials, in which larger values of $\left| B^{0}_{2} \right| $ favor higher transition temperatures, as observed in CeAu$_{0.92}$Bi$_{1.6}$ \cite{CrisAu}, CeCuBi$_{2}$ \cite{CrisCu}, CeNi$_{1-x}$Bi$_{2}$ \cite{CeNiBi}, CeCd$_{1-\delta}$Sb$_{2}$ \cite{CeCdSb2} and UAuBi$_{2}$ \cite{UAuBi2}.

\begin{figure}[!t]
	\includegraphics[width=0.5\textwidth]{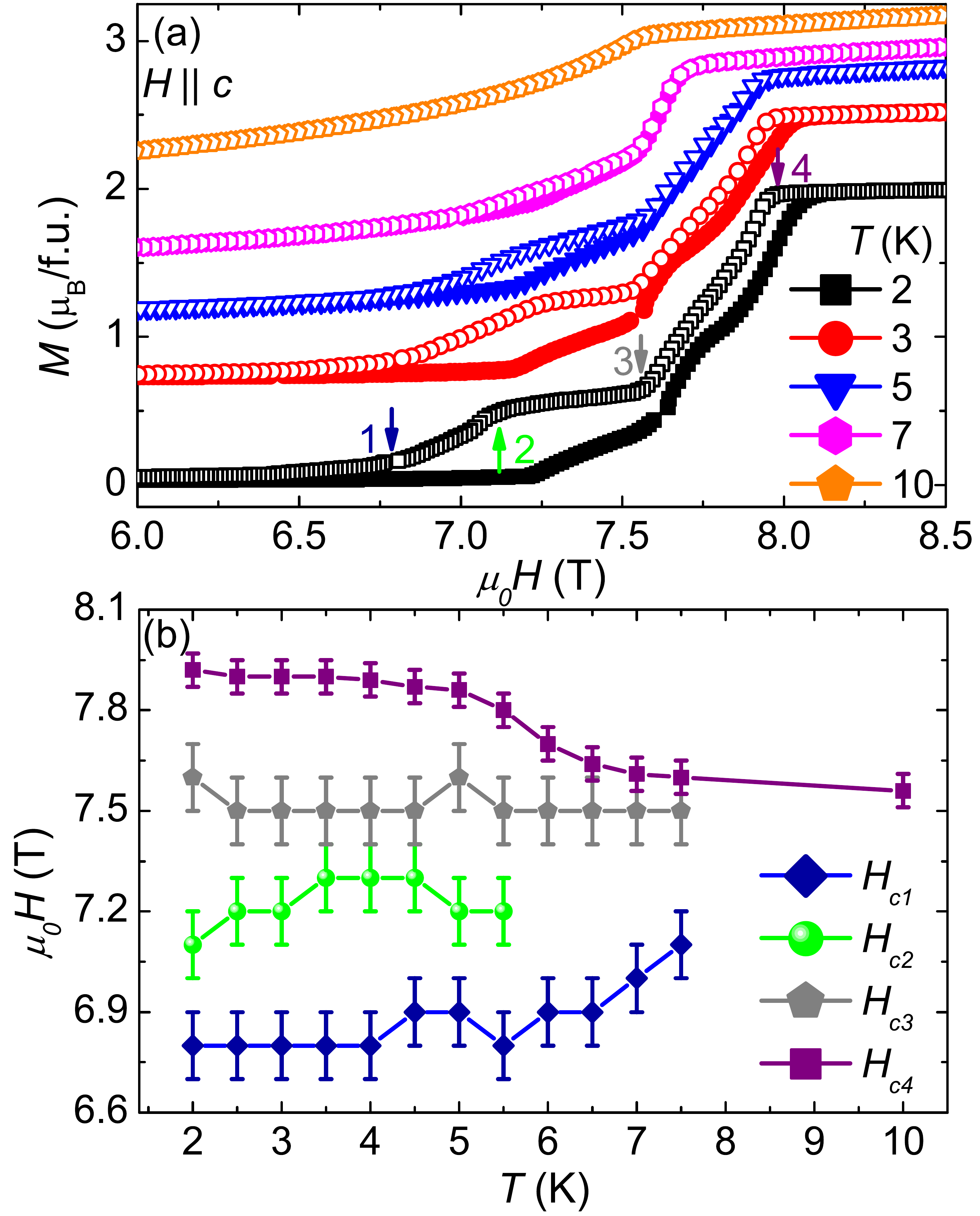}
	\caption{ (a) Magnetization as a function of applied magnetic fields at different temperatures for fields parallel to the $c$-axis. Solid symbols represent data taken with increasing magnetic field and open symbols with decreasing fields. (b) Critical fields as a function of temperature.}
	\label{MxHxT}
\end{figure}

\begin{figure}[!t]
	\includegraphics[width=0.5\textwidth]{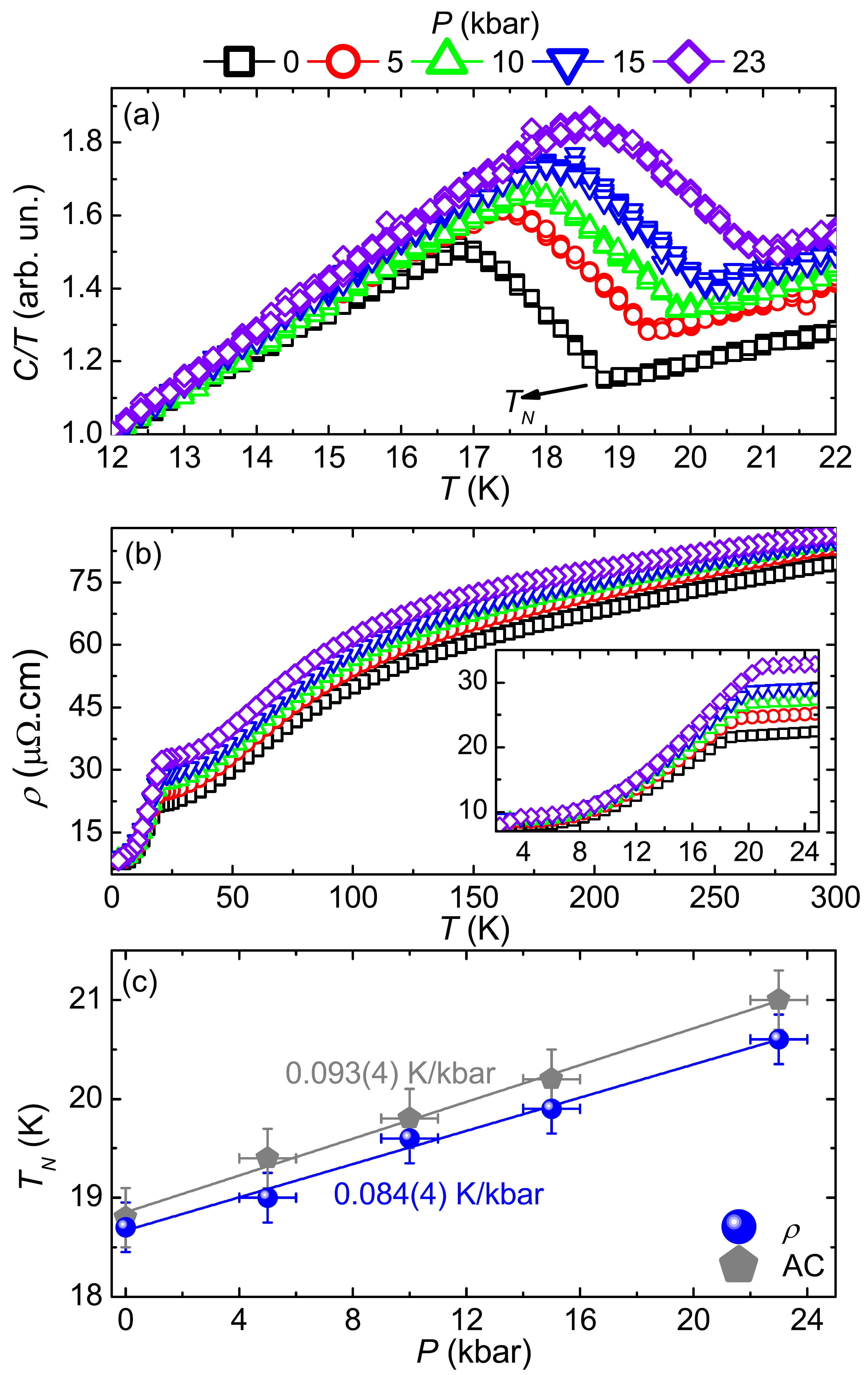}
	\caption{(a) AC calorimetry as a function of temperature at several pressures. (b) Electrical resistivity as a function of temperature at several pressures. (c) Temperature-pressure phase diagram for CeAuBi$_{2}$.}
	\label{AC_rho}
\end{figure}

Figure~\ref{MxHxT} presents the magnetization as a function of applied field at different temperatures for fields parallel to the $c$-axis. The curves were vertically shifted to improve visualization. At 2~K and with increasing magnetic fields (solid symbols),  one can see three discontinuities and a smooth change of slope in the magnetization curve. These anomalies may be related to metamagnetic transitions which can be caused by changes in the magnetic structure of the compound. With decreasing field (open symbols) four anomalies are observed and a large hysteresis appears. This region decreases with increasing temperature. Figure~\ref{MxHxT}(b) summarizes the temperature evolution of the critical fields defined as indicated by the arrows in Figure~\ref{MxHxT}(a). One can clearly see the suppression of the hysteresis as a function of temperature. At 10~K, only one transition is visible. The presence of these spin transitions is an indication that this compound may present magnetic structure transitions as a function of applied magnetic field, as occurs in CeAuSb$_{2}$ \cite{Phase, Stress, Multiq, SSeo}, CeAgBi$_{2}$ \cite{SMT} and CeSbTe \cite{CeSbTe}. Future field dependent neutron diffraction experiments will be valuable to explore this possibility.  

\begin{figure*}[!t]
	\includegraphics[width=\textwidth]{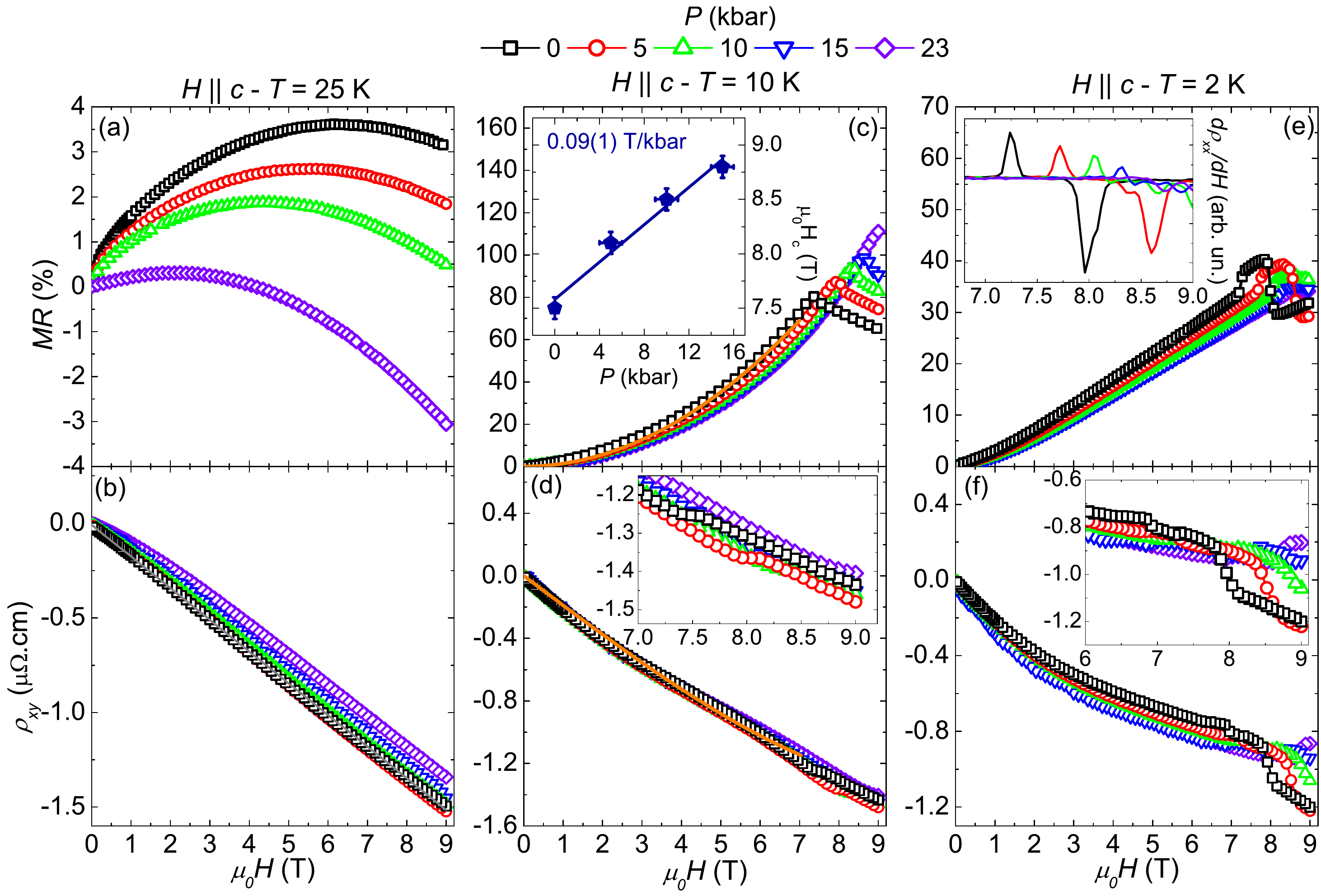}
	\caption{Magnetoresistance ($MR$) and Hall resistivity ($\rho_{xy}$) as a function of applied magnetic field for several pressures at three different temperatures. The solid gray and orange lines are one band and two-band model fits, respectively.}
	\label{Hall_MR}
\end{figure*} 

Figure~\ref{AC_rho} (a) shows the AC heat capacity divided by the temperature as a function of temperature for different external pressures. At ambient pressure, the antiferromagnetic transition occurs at 19~K, in excellent agreement with the specific heat data. One can clearly see the enhancement of $T_{N}$ with increasing pressure reaching 21~K at 23~kbar. Figure~\ref{AC_rho} (b) presents the electrical resistivity as a function of temperature for several pressures. At high temperatures, CeAuBi$_{2}$ displays a metallic behavior in the entire pressure range studied. The inset of Fig.~\ref{AC_rho} (b) displays a closer view of the low temperature behavior of the electrical resistivity as a function of temperature and pressure. A kink in the resistivity reveals the onset of magnetic ordering at 19~K at ambient pressure, which reaches 21~K at 23~kbar. To summarize the evolution of the antiferromagnetism in CeAuBi$_{2}$ as a function of pressure, we show the temperature-pressure phase diagram displayed in Fig.~\ref{AC_rho} (c). This phase diagram demonstrates the enhancement of $T_{N}$ as a function of pressure with a slope of 0.093(4)~K/kbar and 0.084(4)~K/kbar, extracted from the AC heat capacity and electrical resistivity measurements, respectively. In contrast to CeCuBi$_{2}$, in which $T_{N}$ is slowly suppressed as a function of pressure \cite{MarioCu}, the application of external pressure in CeAuBi$_{2}$ enhances the antiferromagnetism. This opposite behavior is consistent with the Doniach's diagram and the fact that the Cu atom is smaller than Au. In this regard, for CeAuBi$_{2}$ the application of external pressure still favors the RKKY interaction instead of the Kondo effect, leading to higher antiferromagnetic transition temperatures. However, for CeCuBi$_{2}$, the Kondo effect is enhanced by external pressure, suppressing $T_{N}$.   

Figure~\ref{Hall_MR} displays MR and the Hall resistivity ($\rho_{xy}$) of CeAuBi$_{2}$ as a function of applied magnetic field ($H \; || \; c$) for several pressures at three different temperatures. At 25~K (Fig.~\ref{Hall_MR}(a)), a negative concavity is observed in MR at all pressures studied. Increasing pressure favors the negative response of MR as a function of field. This behavior may be associated with a spin dependent scattering mechanism, as $T_{N}$ is increasing with pressure, thus short-range interactions, which favor negative MR, are higher at 25~K and 23~kbar than at 25~K and ambient pressure. The negative Hall resistivity indicates that the transport properties in this system are dominated by electrons, as presented in Fig.~\ref{Hall_MR}(b). At 25~K and ambient pressure, $\rho_{xy}$ linearly decreases as a function of magnetic field. A carrier density of $n_{e} = 3.8(1) \times 10^{21}$~e/cm$^{3}$ can be extracted for all studied pressures by performing linear fits, considering $R_{H} = 1/(en)$, in which $e$ is the electron charge.  Increasing pressure slowly enhances a nonlinear response of $\rho_{xy}$ and at 23~kbar a small curvature can be seen for small fields. At 10~K and ambient pressure (Fig.~\ref{Hall_MR} (c)), the MR increases with increasing field and reaches 80~\% at 7.5~T, when a spin-flop transition takes place, in agreement with the magnetization data. For higher fields the MR decreases as a function of field. The MR and the critical field are enhanced with increasing pressure; the first reaches 110~\% at 23~kbar, while the second becomes larger than 9~T at 23~kbar. The inset of Fig.~\ref{Hall_MR}(c) clearly shows the evolution of $H_{c}$ as a function of pressure, which increases at 0.09(1)~T/kbar, similar to the increasing rate of $T_{N}$, if $k_{B}T_{N} = g \mu_{B} H$, where $g = 6/7$. The Hall response as a function of field at 10~K and ambient pressure is again linear and negative, as presented in Fig.~\ref{Hall_MR}(d). Moreover, the magnetic transition observed in MR and magnetization measurements also leads to a discontinuity in $\rho_{xy}$, better seen in the inset of Fig.~\ref{Hall_MR}(d). This discontinuity occurs at higher fields as pressure is increased, in agreement with MR measurements. Furthermore, the orange solid line in panels (c) and (d) of Fig.~\ref{Hall_MR} is a representative fit considering a two-band model:  

\begin{eqnarray}\label{2band}
\nonumber \rho_{xx} (H) &=& \frac{1}{e}\frac{(n_{h}\mu_{h} + n_{e}\mu_{e}) + (n_{h}\mu_{e} + n_{e}\mu_{h})\mu_{e}\mu_{h}H^2}{(n_{h}\mu_{h} + n_{e}\mu_{e})^{2}+[(n_{h} - n_{e})\mu_{e}\mu_{h} H]^2} \\
\rho_{xy} (H) &=& \frac{H}{e}\frac{(n_{h}\mu_{h}^2 - n_{e}\mu_{e}^2) + (n_{h} - n_{e})\mu_{e}^2\mu_{h}^2 H^2}{(n_{h}\mu_{h} + n_{e}\mu_{e})^{2}+[(n_{h} - n_{e})\mu_{e}\mu_{h} H]^2} 
\end{eqnarray}

\noindent where $n$ and $\mu$ are the carrier density and the mobility, respectively, for holes ($h$) and electrons ($e$). These fits result in $n_{h} = 2.52(1) \times 10^{20}$~h/cm$^{3}$, $\mu_{h} = 1.08(1) \times 10^{3}$~cm$^{2}$/Vs, $n_{e} = 2.38(1) \times 10^{20}$~e/cm$^{3}$, $\mu_{e} = 1.29(1) \times 10^{3}$~cm$^{2}$/Vs at ambient pressure. These parameters slowly change with pressure, reaching $n_{h} = 2.49(1) \times 10^{20}$~h/cm$^{3}$, $\mu_{h} = 0.97(1) \times 10^{3}$~cm$^{2}$/Vs, $n_{e} = 2.25(1) \times 10^{20}$~e/cm$^{3}$, $\mu_{e} = 1.19(1) \times 10^{3}$~cm$^{2}$/Vs at 23~kbar. It is worth mentioning that these fits may not be unique and should be taken with caution, due to the absence of experimental constraints on the carrier densities and mobilities. Nevertheless, at 10~K the estimated carrier densities are similar to ZrSiS \cite{ZrSiS density}, however the mobilities are one order of magnitude smaller in CeAuBi$_{2}$. This may be an indication that the band crossings present in CeAuBi$_{2}$ are far away from the Fermi level and that the transport properties are dominated by trivial bands.  At 2~K the MR displays a linear response with increasing magnetic fields for all studied pressures, as can be seen in Fig.~\ref{Hall_MR} (e). A linear response of the MR could arise from band crossings near the Fermi level. At high magnetic fields ($\mu_{0}H > 7$~T) and ambient pressure, two metamagnetic transitions can be seen, the first one at 7.2~T and the second at 8.0~T, as better visualized in the inset of Fig.~\ref{Hall_MR} (e), which present the derivative of MR as a function of applied magnetic fields. Increasing pressure enhances both metamagnetic transitions at first, reaching 7.7~T and 8.6~T at 5~kbar. Further increasing pressure may suppress those transitions or move them to fields higher than 9~T, as they cannot be identified anymore to 9~T. Moreover at 9~T and 2~K the MR stays around 35~\% for all studied pressures. Figure~\ref{Hall_MR}(f) presents $\rho_{xy}$ at 2~K for several pressures as a function of applied magnetic field. The nonlinear response of the Hall resistivity indicates that multi-band effects are present at this temperature for CeAuBi$_{2}$. Moreover, three anomalies can be seen at high magnetic fields, better displayed in the inset of Fig.~\ref{Hall_MR} (f), which are enhanced with increasing pressure.

For temperatures above $T_{N}$, neutron diffraction experiments found Bragg peaks at positions consistent with the $P4/nmm$ space group. Below $T_{N}$, a new set of peaks emerged at positions ($h$, $k$, $(2n+1)/2$), where $h$, $k$ and $n$ are integers. These peaks are consistent with an antiferromagnetic structure with propagation vector $\hat{\tau} = [0, 0, 1/2]$, which doubles the paramagnetic unit cell along the $c$-axis.

\begin{table}[!b]
	\begin{center}
		\caption{Integrated magnetic Bragg intensities and calculated magnetic intensities for different types of magnetic structures. The intensities are in arbitrary units.}
		\label{intn}
		\begin{tabular}{c c c c c c c}
			\hline
			& &  Model & $++--$ & Model & $+--+$ & \\
			(1, 1, $l$) &    $I_{obs}$ &  $\hat{\boldsymbol{\eta}}||c$ & $\hat{\boldsymbol{\eta}}||ab$ & $\hat{\boldsymbol{\eta}}||c$ & $\hat{\boldsymbol{\eta}}||ab$ &   \\
			\hline
			1/2   & 100(2) &    100.0  &  69.5  & 100.0 & 79.1 &  \\
			3/2  &  100(1)  &   96.0 &  94.3  & 53.4 & 59.7 &  \\
			5/2  &   28.3(5) &  31.0 &  48.2  & 56.5 & 100 &   \\
			7/2  &  30.9(2)  & 41.4 &  100.0  & 12.3 & 33.8 &   \\
			9/2  &   5.00(2) &  6.4  &   22.7  & 22.7 & 92.1 &  \\
			11/2  &   16.8(6) &  17.5 &  87.3  & 2.5 & 14.1 & \\
			\hline
		\end{tabular} 
	\end{center}
\end{table}

The magnetic Bragg intensity is defined as \cite{JLynn}:

\begin{equation}\label{neutron}
I_{M}(q) = N_{q} \left( \frac{\gamma r_{0}}{2} \right)^{2} \left| F_{M}(q) \right| ^{2}, 
\end{equation}

\noindent where $I_{M}$ is the integrated intensity for the magnetic reflection, $q$ is the reciprocal lattice vector, $N_{q}$ is a constant that depends on the experimental details, $\left( \frac{\gamma r_{0}}{2} \right)^{2}$ is the neutron-electron coupling constant (0.07265~b/$\mu_{B}^2$), and  $F_{M}$ is the magnetic structure factor.

Let us assume that the magnetic structure of CeAuBi$_{2}$ is similar to CeCuBi$_{2}$ \cite{CrisCu}. In this case the magnetic structure is collinear and the magnetic moments are aligned along a unique direction of the structure (the $c$-axis). Therefore, the magnetic structure factor can be simplified to: 

\begin{equation}\label{magfactor}
\left| F_{M}(q) \right| ^{2} = \left\langle 1 - \left(  \hat{\boldsymbol{q}} \cdot \hat{\boldsymbol{\eta}}  \right)^{2}  \right\rangle \left\langle M \right\rangle^{2} f^{2}(q) \left| \sum_{j} \eta_{j} e^{i \boldsymbol{q} \cdot \boldsymbol{r}_{j}}  \right|^{2}   , 
\end{equation}

\noindent in which $\hat{\boldsymbol{\eta}}$ is the direction of the ordered moment, $\left\langle M \right\rangle$ is the average value of the ordered moment, $f(q)$ is the Ce$^{3+}$ magnetic form factor \cite{fatorforma}, $\eta_{j}$ is the sign of the magnetic moment ($+1$ or $-1$) and $\boldsymbol{r}_{j}$ is the position of the magnetic ions in the unit cell. Equations~\ref{neutron} and \ref{magfactor} enable us to calculate the intensities of the magnetic Bragg reflections considering different directions of the Ce$^{3+}$ magnetic moment. To simulate the observed intensities we considered two scenarios, with a ferromagnetic ($++--$) or antiferromagnetic ($+--+$) coupling between the Ce$^{3+}$ ions within a unit cell. Furthermore, we also considered that the Ce$^{3+}$ magnetic moment could be parallel to the $a$, $b$ or $c$-axis. Table~\ref{intn} summarizes these simulations along with the integrated magnetic intensities. We note that canting of the spins was not considered in these simulations. Nevertheless, one can clearly see that the best model is the $++--$ with the Ce$^{3+}$ parallel to the $c$-axis. Therefore the magnetic structure of CeAuBi$_{2}$ is the same as CeCuBi$_{2}$ \cite{CrisCu}. However, its dependence with applied magnetic field could be rather complex, as revealed by the magnetization measurements. In this regard,  microscopic experiments, such as future neutron diffraction, as a function of applied magnetic field would be interesting to probe the evolution of the magnetic structure of CeAuBi$_{2}$ with field.

\begin{figure}[!t]
	\includegraphics[width=0.5\textwidth]{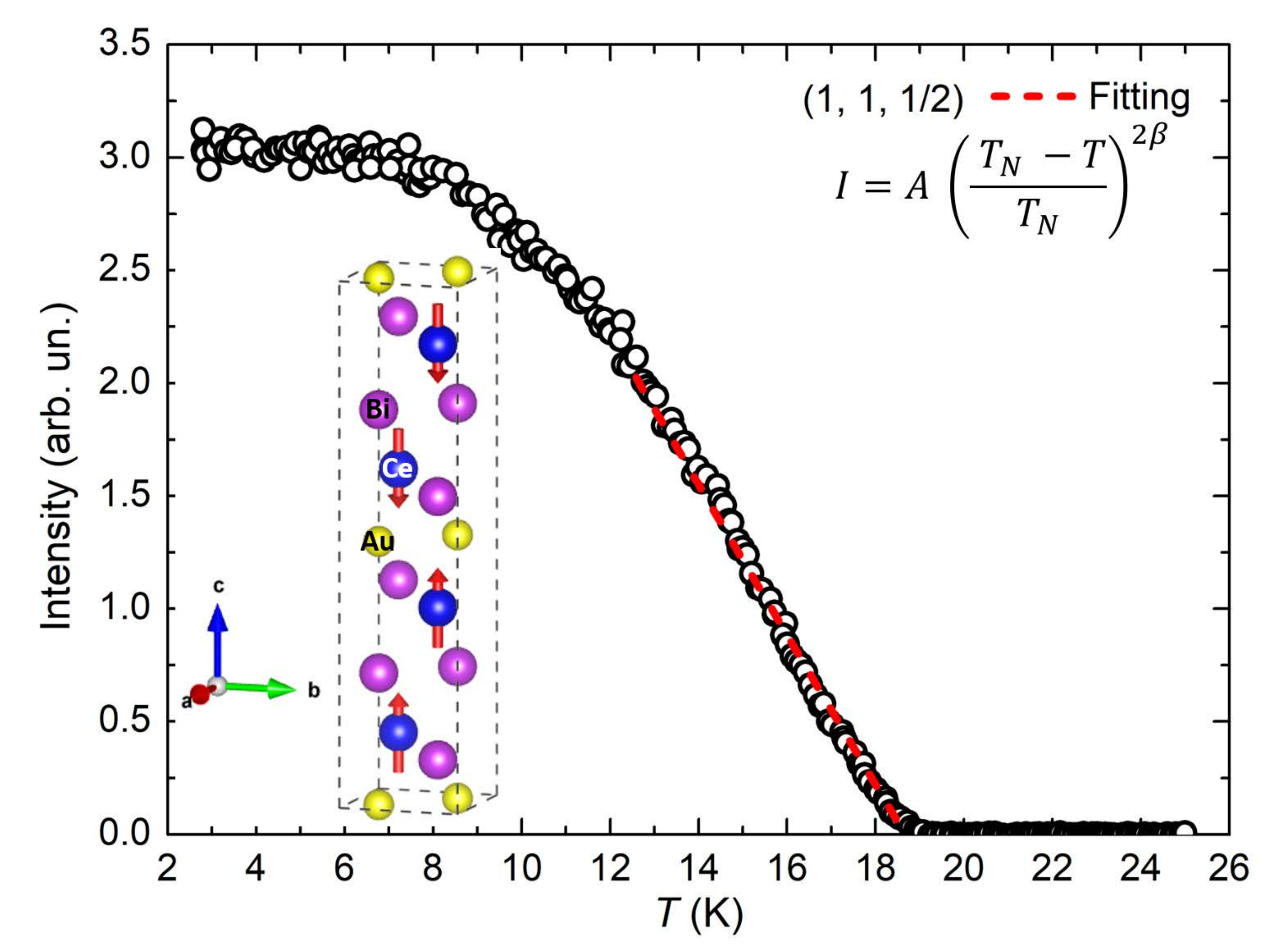}
	\caption{(a) (1, 1, 1/2) reflection intensity as a function of temperature. The dashed red line is a mean field fit. The inset shows a schematic representation of the magnetic structure of CeAuBi$_{2}$. The error bars are smaller than the data points.}
	\label{MOP}
\end{figure} 

Finally, by measuring nuclear Bragg reflections at (1, 1, 0), (1, 1, 1), (1, 1, 2), (1, 1, 3), (1, 1, 4) and (1, 1, 5), we estimated $N_{q}$ to extract the average value of the ordered moment ($\left\langle M \right\rangle$) per Ce ion. The magnetic Bragg reflections presented in Table~\ref{intn} yield an average ordered moment of 2.3(4)~$\mu_{B}$ per Ce$^{3+}$ ion for the ground state ordered moment, in agreement with our magnetization results. Figure~\ref{MOP} presents the temperature evolution of the peak intensity for the (1, 1, 1/2) reflection. As expected, the intensity increases below $T_{N} \approx 19$~K. Moreover, the evolution of the intensity as a function of temperature is well fit by a model  $ I = A \left(  \frac{T_{N}-T}{T_{N}} \right) ^{2\beta}$, for temperatures near $T_{N}$. This fit results in a $T_{N} \approx 19$~K, in agreement with other measurements and a $\beta$ of $0.50(1)$ identical to the mean field value of $1/2$, which supports a localized scenario for the Ce$^{3+}$ ions similar to CeCuBi$_{2}$ \cite{CrisCu}. The inset of Fig.~\ref{MOP} displays a schematic representation of the magnetic structure of CeAuBi$_{2}$. 

We remark that  electronic structure calculations of CeAuBi$_{2}$ support this magnetic structure. We chose the magnetic moment pointing along the $c$-axis and considered three magnetic structures (A)$++--$, (B) $+--+$ and (C) $++++$. For all these three structures, we obtained consistently a spin moment of $\mu_S = -0.966$~$\mu_B$ and orbital moment of $\mu_L = 2.916$~$ \mu_B$. The opposite sign of the spin and orbital moments is expected from the first Hund´s rule. The resultant total magnetic moment of about $1.95$~$\mu_B$ agrees well with the magnetization saturation shown in Fig.~\ref{chi} (d) and with the ordered moment extracted from the neutron diffraction experiments.  In addition, we found that the magnetic configuration (A) has the lowest total energy as a ground state, which is also consistent with neutron measurements.  One can fit the total energy results for all three configurations to a Heisenberg model, and the results indicate the presence of two distinct exchange interactions $J_{\parallel,1}^{ex} = -0.9 $~meV and $J_{\parallel,2}^{ex} =0.5$~meV, in qualitative agreement with our CEF models.

\section{CONCLUSIONS}

In summary,  we present the electronic and magnetic properties of stoichiometric CeAuBi$_{2}$. At room temperature, CeAuBi$_{2}$ crystallizes in the $P4/nmm$ structure with lattice parameters $a = $4.628(6)~\AA \- and $c = $9.897(13)~\AA. At ambient pressure, it orders antiferromagnetically at $T_{N} = 19$~K.  Furthermore, by performing fits of $\chi(T)$ and $M(H)$ using a CEF mean field model, we could extract two competing exchange interactions, $z_{AFM}J_{AFM}^{ex} = 1.65$~K and $z_{FM}J_{FM}^{ex} = -0.35$~K and a $\Gamma_{7}^{1} = 0.99|\pm5/2\rangle - 0.08|\mp3/2\rangle$ ground state. Several  metamagnetic transitions at 2~K with fields parallel to the $c$-axis are present in CeAuBi$_{2}$. These transitions indicate that the magnetic structure changes as a function of applied magnetic fields. Therefore, microscopic measurements, such as neutron diffraction as a function of applied magnetic field, would be helpful to shed light on this issue. In contrast to CeCuBi$_{2}$ \cite{MarioCu}, the application of external pressure in CeAuBi$_{2}$ enhances $T_{N}$ to 21~K at 23~kbar. Experiments under higher pressures need to be done in CeAuBi$_{2}$ to probe if $T_{N}$ can be suppressed, inducing a quantum critical point. Moreover, MR and Hall experiments enabled the estimate of $n_{h} = 2.52(1) \times 10^{20}$~h/cm$^{3}$, $\mu_{h} = 1.08(1) \times 10^{3}$~cm$^{2}$/Vs, $n_{e} = 2.38(1) \times 10^{20}$~e/cm$^{3}$, $\mu_{e} = 1.29(1) \times 10^{3}$~cm$^{2}$/Vs at 10~K and ambient pressure. The application of external pressure does not affect these parameters significantly. Moreover, these carrier densities are similar to the ones found in ZrSiS \cite{ZrSiS density}, however the mobilities are one order of magnitude smaller in CeAuBi$_{2}$, which indicates that trivial bands dominate the transport properties of this compound. At 2~K a linear response of MR when fields are applied parallel to the $c$-axis was observed. Experiments at higher magnetic fields are in need to explore the evolution of this unusual behavior. Finally, neutron magnetic diffraction experiments revealed an antiferromagnetic propagation vector $\hat{\tau} = [0, 0, 1/2]$, which doubles the paramagnetic unit cell along the $c$-axis. The magnetic structure presents a ferromagnetic coupling between the Ce$^{3+}$ ions within the unit cell ($++--$), identical to CeCuBi$_{2}$ \cite{CrisCu}. At 2.8~K, the average magnetic moment reaches 2.3(4)~$\mu_{B}$ per Ce$^{3+}$ ion.

\begin{acknowledgments}
This work was supported by the S\~ao Paulo Research Foundation (FAPESP) grants 2015/15665-3, 2017/10581-1, 2017/25269-3, 2018/11364-7, 2019/04196-3, CNPq grant $\#$ 304496/2017-0 and CAPES, Brazil. The experimental and theoretical work at Los Alamos was performed under the auspices of the U.S. Department of Energy, Office of Basic Energy Sciences, Division of Materials Science and Engineering under projects ``Quantum Fluctuations in Narrow-Band Systems'' and ``Integrated Modeling of Novel and Dirac Materials'', respectively. The identification of any commercial product or trade name does not imply endorsement or recommendation by the National Institute of Standards and Technology.
\end{acknowledgments}

\bibliography{basename of .bib file}

\end{document}